# Plasmonic Metasurfaces for Magnetic Skyrmion Control via the Inverse Faraday Effect

Xingyu Yang[1,2°], Tristan da Câmara Santa Clara Gomes[3,4°], Chantal Hareau[1], Ye Mou[5], Tingjun Zheng[6,7], Maria Sanz-Paz[1], Vincent Cros[8], Nicolas Reyren[8], Jack Gartside[6,7] and Mathieu Mivelle[1, *]

[1]Sorbonne Université, CNRS, Institut des NanoSciences de Paris, INSP, F-75005 Paris, France

[2]School of Optoelectronic Engineering and Instrumentation Science, Dalian University of Technology, Dalian 116024, China.

[3]Instituto de Engenharia de Sistemas E Computadores – Microsistemas e Nanotecnologias (INESC MN), 100-029 Lisbon, Portugal

[4]Institute of Condensed Matter and Nanosciences, Université Catholique de Louvain, 1348 Louvain-la-Neuve, Belgium

[5]School of Electronic and Information Engineering, Ningbo University of Technology, No. 201, Fenghua Road, Jiangbei District, Ningbo, Zhejiang, China

[6]Blackett Laboratory, Department of Physics, Imperial College London, London SW7 2AZ, United Kingdom

[7]London Centre for Nanotechnology, Imperial College London, London SW7 2AZ, United Kingdom

[8]Laboratoire Albert Fert, CNRS, Thales, Université Paris-Saclay, 91767 Palaiseau, France

° Equal contribution

*Corresponding authors:

mathieu.mivelle@sorbonne-universite.fr

## Abstract

Magnetic skyrmions hold high promise for low-power spintronic and neuromorphic computing. However, existing manipulation schemes rely predominantly on local thermal excitation, rendering nucleation stochastic and lacking a non-destructive mechanism for targeted erasure. Here, we demonstrate a deterministic, optomagnetic field-driven paradigm for fast, all-optical writing and erasing of magnetic skyrmion crystals using a magneto-plasmonic metasurface. By tailoring surface lattice resonances in a periodic nanodisk array, circularly polarized light excites giant super-circular optical spin densities that drive intense circulating drift photocurrents via the inverse Faraday effect. This delivers synchronized picosecond magnetic field pulses into an adjacent magnetic multilayer. Micromagnetic simulations reveal that a single optical pulse deterministically nucleates stable Néel skyrmions through a transient Bloch-to-Néel relaxation pathway governed by optical helicity and interfacial Dzyaloshinskii–Moriya interaction. Crucially, reversing incident light helicity allows on-demand topological reconfiguration: under a confining bias field, the inverted optomagnetic field unwinds pre-existing skyrmions to restore a uniform ground state. Furthermore, we demonstrate the targeted writing of stable skyrmionium states. By bypassing stochastic thermal cycles and achieving reversible topological control on picosecond timescales, this work bridges nanophotonics, plasmonics, and magnetism, establishing a scalable foundation for high-speed, reconfigurable information processing and unconventional computing.

**Keywords:** skyrmion crystal, plasmonic metasurface, inverse Faraday effect, nanophotonics, light-matter interactions, magneto-plasmonics

## 1. **Introduction**

Over the last decade, condensed matter physics has uncovered a wealth of intriguing fundamental phenomena hosted by topological defects,[1,2,3,4,5] with increasing demonstration that their benefits extend into applied functional schemes—notably in high-density data storage and unconventional computing. Particularly fruitful is the class of nanoscale topological structures termed magnetic skyrmions,[6,7] and related higher-order textures such as skyrmionium.[8] Originally proposed by Tony Skyrme in nuclear physics,[9] magnetic skyrmions manifest as localized, topologically-protected spin textures characterized by a continuous winding of magnetic moments and a quantized unit topological charge, which can be stabilized in metallic multilayers.[10,11] Their particle-like behavior, nanoscale dimensions, and low-current-driven mobility at low currents make them prime candidates for next-generation spintronic memory, logic, and neuromorphic or physical learning network hardware.[12,13,14,15,16,17,18,19]

While the manipulation and transport of existing magnetic skyrmions via electrical currents, spin waves, or magnetic fields have been extensively explored, methods for their targeted on-demand writing and erasing remain in crucial need of further development.[20,21,22] Conventional optical[23] and electrical[8,19] skyrmion generation approaches predominantly rely on local heating cycles[24] or ultrafast thermal demagnetization.[25,26] These thermal mechanisms are inherently stochastic—failing to reliably control the exact number, position, or chiral state of the nucleated quasiparticles—and typically require structural defects, sample edges, or strong external magnetic fields to nucleate. Purely thermal schemes lack a mechanism to selectively "erase" individual skyrmions or skyrmion lattices without macroscopically perturbing or damaging the surrounding magnetic state. These existing protocols generally lack the ability to selectively control the topological charge of the generated spin textures, leaving on-demand selective writing of distinct topological quasiparticles, such as skyrmions and higher-order skyrmioniums, largely out of reach.[27] Achieving fast, deterministic, non-thermal, and reconfigurable topological control across multiple charge states thus remains a pivotal open challenge.

A promising alternative lies at the intersection of optics[28, 29] and magnetism[30] through the inverse Faraday effect (IFE).[31, 32, 33, 34, 35] The IFE is a magneto-optic phenomenon where circularly polarized light drives non-dissipative circulating drift photocurrents in metallic and plasmonic media, generating magnetic fields on picosecond timescales.[36] Recent advances in nanophotonics and plasmonics have enabled precise spatial engineering of optical spin density[37, 38] and electric-magnetic field gradients at the nanoscale, broadening applications of the inverse Faraday effect by enabling non-conventional regimes such as linearly polarized optomagnetism, chiral inverse Faraday responses, and the spatial tilting of magnetization vectors using optical orbital angular momentum.[39, 40, 41, 42, 43, 44, 45, 46, 47, 48, 49, 50, 51, 52, 53] In particular, periodic plasmonic metasurfaces supporting surface lattice resonances (SLRs), i.e. hybrid modes arising from the coherent coupling between localized surface plasmons and in-plane diffractive orders, can drastically enhance local optical fields and resulting magnetic IFE fields while maintaining low optical loss across extended arrays.

Here, we demonstrate a magneto-plasmonic IFE platform capable of deterministic all-optical writing and erasing of both skyrmions and higher-order skyrmionium in Pt|Co|Al multilayers at picosecond timescales. We design a hexagonal metasurface of Ag nanodisks tailored to support a surface lattice resonance. Upon circularly polarized optical excitation, the metasurface drives intense, circulating drift photocurrents around each nanodisk, delivering picosecond magnetic field pulses of several hundreds of mT into an adjacent magnetic medium, driving deterministic nucleation of topological magnetic textures.

Through micromagnetic simulations (MuMax3) using realistic experimentally-measured parameters of a $[\mathrm{Pt}(3)/\mathrm{Co}(1.3)/\mathrm{Al}(3)]_5$ multilayer with interfacial Dzyaloshinskii–Moriya interaction (DMI), we show that a single 20 ps laser pulse deterministically imprints a stable clockwise (CW) Néel skyrmion crystal without requiring an external magnetic field. The nucleation dynamics follow a transient Bloch-to-Néel relaxation pathway dictated by the optical pump pulse helicity and the material's intrinsic DMI. Remarkably, by engineering the chirality and polarity mismatch between the optomagnetic field and the magnetic stack, we demonstrate that a single optical pulse can also directly nucleate stable skyrmionium states (double winding leading to non topologcial stable structure). Furthermore, we reveal that subsequent helicity-reversed optical pulses incident on written skyrmions unlock reconfigurable topological control: under a confining external perpendicular bias field, the inverted optomagnetic torque deterministically unwinds pre-existing skyrmions to reset the film to its uniform magnetic state, whereas in the absence of an external magnetic bias field, this second pulse drives an alternative pathway to convert the written skyrmion into a stable skyrmionium. By systematically exploring pulse fluence, thermal effects, field landscapes, and multilayer film stack DMI-favoured magnetic chiralities, we establish selection rules for fast, optically-driven IFE topological writing, erasure, and higher-order texture generation, providing a scalable route toward high-speed, reconfigurable topological spintronic and neuromorphic devices.

## 2. Results:

The overall concept of all-optical topological texture writing via magneto-plasmonic metasurface is illustrated in **Figure 1**. **Figure 1a** shows a Ag nanodisk acting as a plasmonic antenna, fabricated on a glass substrate, in direct contact with an adjacent magnetic multilayer stack. The nanodisk is excited by a picosecond-scale circularly-polarized optical pump pulse through the glass substrate, inducing non-dissipative circulating drift photocurrents in the Ag nanodisk via the inverse Faraday effect (IFE).[36] The resulting localized, out-of-plane magnetic field pulse acts directly on the magnetic layer to nucleate a stable skyrmionic spin texture.

To achieve large-scale and reconfigurable control, these plasmonic nanodisk elements are arranged into a periodic metasurface with a hexagonal lattice (**Figure 1b**). Within each individual unit cell, the optical field drives an azimuthal drift photocurrent loop inside the nanodisk (**Figure 1b**, right, red arrows). The theoretical description of these drift currents ($\boldsymbol{J}_d$) is governed by the hydrodynamic action of optical field gradients in the metal:

$$\boldsymbol{J}_d = \frac{1}{2en} \mathrm{Re}\left(\left(-\frac{\nabla \cdot (\sigma_\omega \boldsymbol{E})}{i\omega}\right) \cdot (\sigma_\omega \boldsymbol{E})^*\right)$$

where $e$ is the elementary charge ($e < 0$), $n$ represents the electron density at rest, $\sigma_\omega$ is the dynamic conductivity of Ag, and $\boldsymbol{E}$ is the local optical electric field. When excited by circularly polarized light, each nanodisk in the hexagonal lattice acts as a circular current source, generating a localized out-of-plane magnetic field vector at its center (**Figure 1c**). Through the Biot–Savart law, this hexagonal distribution of current loops produces a periodic, structured magnetic field profile centered on each nanodisk across the metasurface (**Figure 1d**).

The geometry parameters of the metasurface are defined in **Figure 1e**. Nanodisks of radius r and thickness t = 10 nm are positioned on a glass substrate with a lattice period *p* (**Figure 1e**, left). The structure is then optically illuminated from the substrate side (negative z-direction), and the resulting IFE-induced magnetic field is evaluated in a plane located 1 nm above the top surface of the Ag nanodisks (**Figure 1e**, right), at the interface with the magnetic film that would be placed just above.

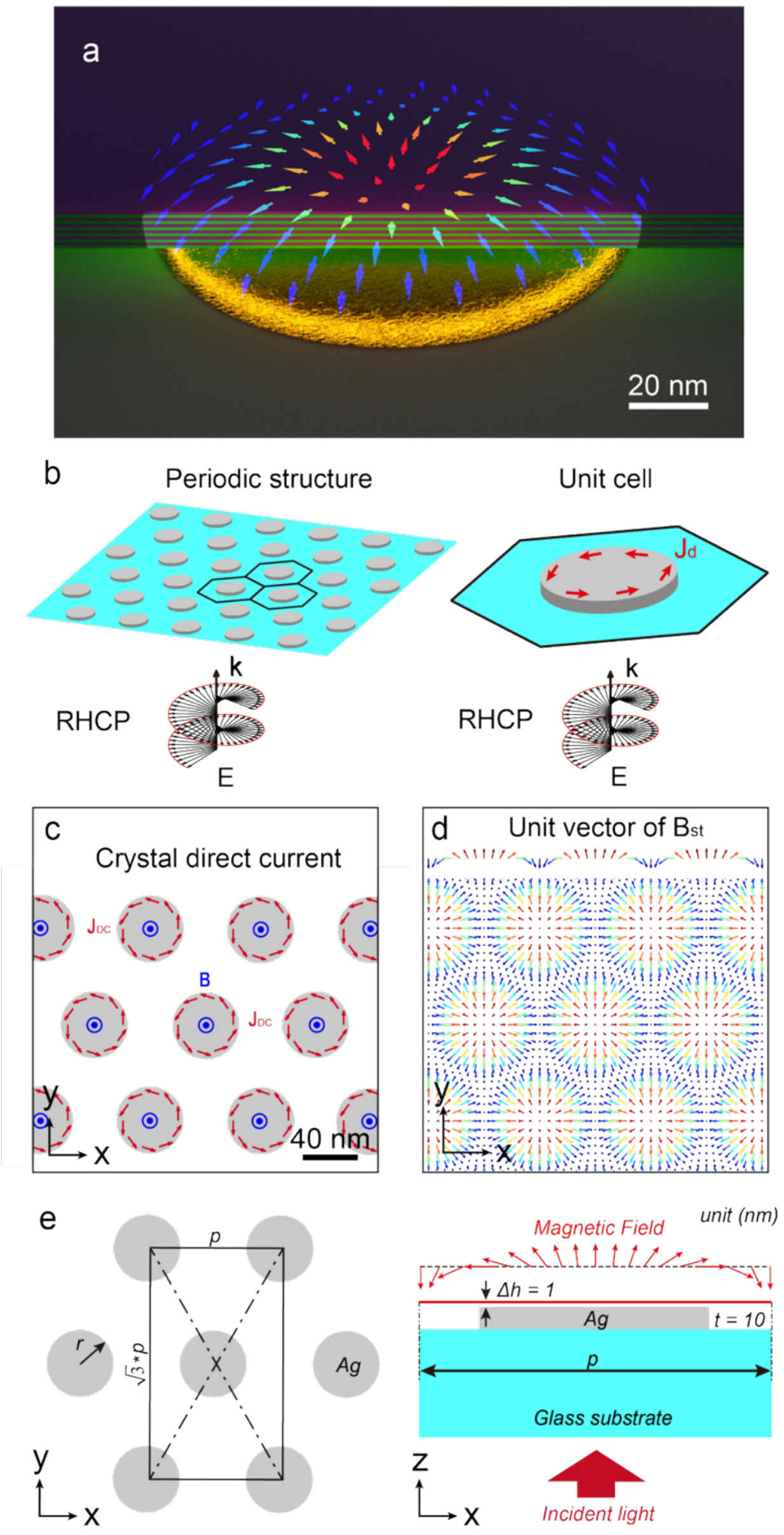

**Figure 1: Working principle of the magneto-plasmonic metasurface. a)** Conceptual illustration of all-optical skyrmion writing: an Ag nanodisk on a glass substrate coupled to a magnetic multilayer generates a localized IFE magnetic field pulse under laser excitation, imprinting a skyrmionic spin texture into the magnetic layer. **b)** Schematic layout of the plasmonic metasurface showing the periodic hexagonal lattice across multiple elements (left) and a single unit cell detailing the optically induced circular drift currents within the nanodisk (red arrows, right). Circular handedness is defined as viewed from the receiver side. **c)** Schematic representation of the circular drift photocurrent loops and their associated out-of-plane magnetic field vectors across the hexagonal metasurface. **d)** Spatial distribution profile of the magnetic field generated by the periodic drift current distribution. **e)** Geometric parameters of the metasurface: top view ($XY$ plane, left) indicating nanodisk radius $r$ and lattice period $p$; cross-sectional view ($XZ$ plane, right) displaying the glass substrate, bottom illumination from the negative $\mathrm{z}$-direction, nanodisk thickness $\mathrm{t} = 10$ nm, and the magnetic field evaluation plane positioned 1 nm above the top Ag interface.

To maximize the magneto-plasmonic response and achieve strong local magnetic fields, we optimize the collective optical response of the metasurface by tailoring its surface lattice resonance (SLR) to match the optical pump pulse (**Figure 2**). The maximum electric field enhancement in the XY plane of the hexagonal metasurface as a function of the lattice period $p$ and the incident excitation wavelength $\lambda$ is shown in **Figure 2a**. A distinct maximum emerges from the diffractive coupling between the localized surface plasmon resonances of individual Ag nanodisks, and the in-plane Rayleigh anomalies.[54] For an Ag nanodisk diameter of 120 nm and a thickness of 10 nm, the strongest enhancement is achieved at an optimal lattice period of $\mathrm{p} = 595\ \mathrm{nm}$ at a resonance wavelen*gth* $\lambda \approx 750\ nm.$

Under this optimized lattice resonance condition, the local electric field amplitude enhancement around an individual nanodisk is mapped in **Figure 2b**, showing a pronounced local intensity enhancement factor of $|\boldsymbol{E}|/|\boldsymbol{E}_0| \approx 90$ concentrated around disk perimeter. When illuminated by circularly polarized light, this enhanced near-field drives a strong local optical spin density (**Figure 2c**):[47, 55]

$$\boldsymbol{s} = \frac{1}{|\boldsymbol{E}_0|^2}\mathrm{Im}(\boldsymbol{E}^* \times \boldsymbol{E})$$

In nanophotonics, the optical spin density $\boldsymbol{s}$ is a vector quantity that characterizes the local polarization state and intrinsic spin angular momentum carried by the electromagnetic field at the nanoscale.[56] In our coordinate system, the out-of-plane component $s_z$ directly maps the local ellipticity of light: positive values ($s_z > 0$) correspond to right-handed elliptical polarization, negative values ($s_z < 0$) indicate left-handed elliptical polarization, and $s_z = 0$ denotes purely linear polarization.

Crucially, because $\boldsymbol{s}$ is normalized by the incident plane-wave intensity $|\boldsymbol{E}_0|^2$, subwavelength near-field confinement and phase gradients around the nanodisk can cause $|s_z|$ to significantly exceed unity ($|s_z| > 1$). This regime corresponds to so-called "super-circular" light, where the local angular momentum density and degree of circularity surpass the theoretical limit of an ideal circularly polarized plane wave in vacuum.[36] As shown in **Figure 2c**, the spatial distribution of $s_z$ exhibits a homogeneous profile around the nanodisk perimeter, reaching an enhanced super-circular value of about $4000$ (with $s_z < 0$, denoting strong left-handed ellipticity).

Because the drift photocurrents underlying the IFE are directly proportional to this optical spin density,[57] this super-circular near-field drives intense circulating drift photocurrents within the Ag nanodisk (**Figure 2d**), computed via the hydrodynamic inverse Faraday effect formulation (Equation 1). These closed-loop drift currents act as stationary current sources on picosecond timescales. Applying the Biot–Savart law to this current distribution yields the localized effective magnetic field displayed in 3D (**Figure 2e**) and in 2D cross-section (**Figure 2f**), evaluated 1 nm above the nanodisk interface.

The spatial distribution of the induced effective magnetic field amplitude, evaluated 1 nm above a single resonant Ag nanodisk, is shown in **Figure 2g**. By harnessing the surface lattice resonance (SLR), the circulating drift photocurrents driven by the inverse Faraday effect yield a strongly concentrated optomagnetic response, producing an intense, localized magnetic field amplitude reaching approximately 0.6 T at the center of the nanodisk for a peak incident laser intensity of 10 GW/cm². Extending this evaluation across multiple unit cells, **Figure 2h** maps the collective optomagnetic landscape across the periodic hexagonal array $(\mathrm{p} = 595\ \mathrm{nm})$. The metasurface establishes a synchronized, hexagonal lattice of subwavelength magnetic hotspots delivering peak magnetic fields of 0.6 T at each nanodisk. This collective resonance directly combines the subwavelength spatial confinement provided via the plasmonic antennae, and the multi-hundred-mT field threshold required for deterministic writing of chiral topological spin textures, enabling writing of both magnetic skyrmions, and higher-order skyrmionium textures.

It should be noted that in our electromagnetic simulations, the Ag metasurface was modeled on a bare glass substrate. This approach deliberately isolates the fundamental optical response to establish the optimal Surface Lattice Resonance (SLR) conditions and the resulting IFE spatial distributions. While the direct physical contact of a metallic magnetic multilayer would predictably damp and slightly red-shift the SLR due to increased optical absorption, the fundamental symmetry of the super-circular near-fields and the spatial profile of the resulting localized magnetic fields remain unchanged. In practical experimental device designs, this damping could be effectively mitigated by replacing the single nanodisk with a Metal-Insulator-Metal (MIM) nano-resonator, comprising two metallic disks separated by a nanometric dielectric spacer, placed directly on top of the magnetic multilayer. This MIM architecture tightly confines the optical mode within the internal dielectric gap, preserving a high-quality resonance and minimizing parasitic absorption from the substrate, while still delivering strong, highly localized effective magnetic fields to the adjacent magnetic stack.[58]

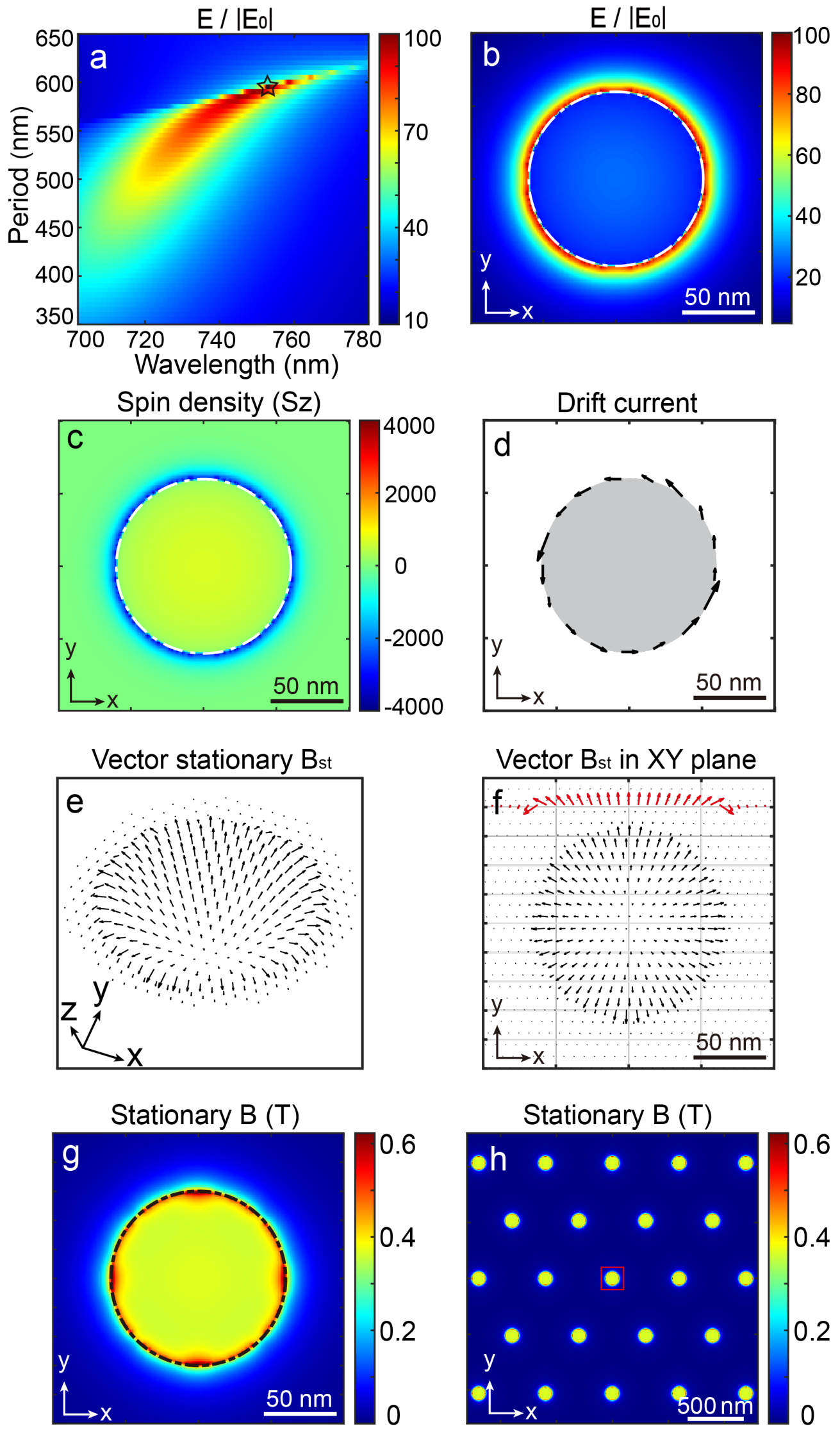

**Figure 2: Surface lattice resonance optimization and spatial distribution of the IFE-induced magnetic field. a)** In-plane electric field enhancement of the hexagonal metasurface as a function of the lattice period $\mathrm{p}$ and incident wavelength $\lambda$, showing the optimum surface lattice resonance (SLR) at $\mathrm{p} = 595\,\mathrm{nm}$ and $\lambda \approx 750\,\mathrm{nm}$. **b)** Spatial distribution of the optical electric field amplitude enhancement ($|\boldsymbol{E}|/|\boldsymbol{E}_0|$) around a single resonant nanodisk in the unit cell. **c)** Spatial map of the normalized out-of-plane optical spin density $s_z$ around the nanodisk under circularly polarized illumination. The amplitude exceeds unity ($|s_z| > 1$), indicating a local "super-circular" polarization regime where the angular momentum density surpasses that of the incident plane wave. **d)** Calculated circular drift photocurrent density ($\boldsymbol{J}_d$) inside the resonant Ag nanodisk. **e)** Three-dimensional vector representation and **f)** two-dimensional color map of the localized magnetic field generated

via the inverse Faraday effect, evaluated 1 nm above the top surface of the nanodisk. **g)** Spatial distribution of the effective magnetic field amplitude evaluated 1 nm above an individual resonant nanodisk at a peak incident intensity of $10$ GW/cm$^2$, reaching a localized peak of approximately 0.6 T (600 mT). **h)** Extended spatial distribution of the magnetic field amplitude across multiple unit cells of the hexagonal array ($\mathrm{p} = 595$ nm), demonstrating a synchronized lattice of subwavelength magnetic hotspots delivering ~0.6 T per site; the local optomagnetic field amplitude scales linearly with the incident laser intensity. [36]

To evaluate the induced magnetic dynamics from the optically-driven magneto-plasmonic IFE fields, we perform micromagnetic simulations of the response of the magnetic multilayer film to the optical IFE fields using MuMax3.[59] We model a $[\mathrm{Pt}(3)/\mathrm{Co}(1.3)/\mathrm{Al}(3)]_5$ multilayer placed in direct contact with the hexagonal Ag nanodisk array, with this film stack having previously demonstrated experimental support for skyrmion textures.[19, 60, 61] The magnetic layer is parameterized using experimentally determined values[62] for exchange stiffness, perpendicular magnetic anisotropy, interfacial DMI and saturation magnetization. The optical excitation propagates along +z from the substrate side, generating a clockwise (CW) optomagnetic field landscape above the nanodisks. In our simulation coordinate system, the magnetic multilayer is represented with its experimental growth sequence inverted along z, such that the top surface of the deposited stack faces the Ag nanodisks. This spatial mapping preserves the high-quality interfaces and magnetic properties of the experimental deposition order while mathematically translating the native stack chirality into a negative interfacial DMI sign in MuMax3, which stabilizes CW Néel skyrmions - matching the spatial profile of the optically-driven IFE field (the usual convention to define the chirality is reversed due to the choice of the z direction from surface to substrate). As the inverse Faraday effect (IFE) is a quasi-instantaneous magneto-optical response, a 20 ps (FWHM) optical pulse with a peak intensity of $10.4\ \mathrm{GW/cm^2}$ generates an effective magnetic field pulse of identical temporal profile to the laser pump pulse (**Figure 3a**).

The dynamic response of the system is tracked via the time evolution of the unit-cell-averaged out-of-plane magnetization ($m_z$), the normalized topological charge ($Q$), and the incident laser pump pulse intensity (**Figure 3a**). The topological charge is calculated using the discrete lattice framework of Berg and Lüscher[63], and normalized per skyrmion (divided by two for the simulated double-skyrmion hexagonal unit cell domain), and averaged over the five magnetic layers, such that (Q=1) corresponds to one skyrmion in each layer. Starting from a uniformly saturated state along the negative direction ($m_{\mathrm{z},init} = -1$), an RHCP laser pulse induces an effective out-of-plane field oriented along $+z$. The magnetic dynamics are evolved over several nanoseconds followed by a final relaxation step, yielding a resultant stable, optically-written clockwise (CW) Néel skyrmion crystal (**Figure 3b**). Symmetrically, applying an LHCP pulse on an initially positive saturated state ($m_{z,init} = +1$) stabilizes a skyrmion lattice with negative core polarity, and identical CW Néel chirality (**Figure 3c**), demonstrating that the IFE field reaches sufficient amplitude to drive deterministic topological nucleation for both uniform $m_{z,init} = \pm 1$ magnetic states.

The microscopic nucleation pathway leading to these stable states is captured in the time-resolved spatial snapshots of the top magnetic layer (**Figures 3d,e**), directly tracking the sequential dynamic stages of magnetic texture evolution (i–v) marked in **Figure 3a**. At point (i) ($0\ \mathrm{ps}$), the system is initialized in a uniformly perpendicularly-magnetized ground state ($m_{z,init} = -1$, $Q = 0$). When the optical pulse reaches its maximum intensity at point (ii) ($40\ \mathrm{ps}$), the spatially non-uniform IFE field generated around the nanodisk initiates local spin reorientation, forming a transient ring-shaped magnetic texture with in-plane magnetization exhibiting a predominantly Bloch-like helicity, dictated by the circular polarization of incident laser pulse. As the pulse amplitude decays at point (iii) ($160\ \mathrm{ps}$), this ring expands and becomes increasingly well-defined while persisting even after the optical excitation has vanished; at this stage, the central core under the nanodisk center has not yet fully reversed. Core reversal develops during the subsequent relaxation, as shown at point (iv) ($190\ \mathrm{ps}$), rather than occurring abruptly at the pulse peak. The individual magnetic layers undergo this topological transition sequentially, producing the stepwise increase of the normalized topological charge observed in **Figure 3a**. At point (v) ($220\ \mathrm{ps}$), the reversed core is then fully established in the upper layer as the multilayer continues to relax. During this post-pulse evolution, interfacial DMI progressively transforms the initially Bloch-like texture into the final stable CW Néel skyrmion.

An identical transient Bloch-to-Néel pathway is observed for symmetric LHCP excitation of the $m_{z,init} = +1$ state (**Figure 3e**). Layer-resolved dynamics (**Supplementary Figures S1–S3**) reveal a stable hybrid 3D texture in which the three upper layers adopt a CW Néel configuration, while the bottom layer develops CCW chirality due to interlayer dipolar coupling, and the intermediate layer stabilizes as a hybrid Bloch–Néel texture matching the azimuthal circulation of the transient state.

The parameter space of this magneto-plasmonic optical writing process is mapped in **Figures 3f–h** as a function of peak laser intensity and laser pulse duration (FWHM). At $\mathrm{T} =$ $0\ \mathrm{K}$, successful skyrmion writing (yellow region) requires higher laser intensities (**Figure 3f**). When thermal fluctuations are incorporated during the dynamic excitation phase, the operational parameter window for successful writing expands significantly at $\mathrm{T} = 300\ \mathrm{K}$ (**Figure 3g**) and $\mathrm{T} = 450\ \mathrm{K}$ (**Figure 3h**), demonstrating that thermal energy assists the IFE field in overcoming the topological nucleation barrier without triggering stochastic multi-domain fragmentation. Symmetrical outcomes are obtained for RHCP excitation on a $-z$-initialized stack and LHCP excitation on a $+z$-initialized stack.

The influence of an external bias magnetic field ($\mathrm{B_{bias}}$) applied perpendicular to the layer is presented in **Figure 3i**. For RHCP excitation of a $-\mathrm{z}$-initialized stack, varying $\mathrm{B_{bias}}$ from $-40\ \mathrm{mT}$ to $+40\ \mathrm{mT}$ decreases the required nucleation energy threshold monotonically from $27\ \mathrm{pJ}$ to $22\ \mathrm{pJ}$. Concurrently, the final relaxed skyrmion diameter exhibits an inverse trend, expanding from $129\ \mathrm{nm}$ up to $557\ \mathrm{nm}$ (approaching the $595\ \mathrm{nm}$ lattice boundary; see **Supplementary Figures S4, S5**). Applying a bias field opposing the skyrmion core ($\mathrm{B_{bias}} < 0$, shaded gray area) increases the energy barrier, but importantly stabilizes compact skyrmions with reduced diameter, whereas a parallel field ($\mathrm{B_{bias}} > 0$) lowers the writing threshold at the expense of core expansion to large diameter, providing a practical knob to balance energy efficiency against spatial density of the written skyrmions.

We then investigate the roles of the chirality and polarity of the IFE field by comparing the nucleation energy threshold for four magnetic configurations under fixed RHCP illumination (**Figure 3j**). Here, the optical excitation remains identical, while the sign of the interfacial DMI and the initial magnetization $m_{\mathrm{z},init}$ are independently varied. Inverting the DMI sign reverses the preferred Néel chirality of the stack (CW vs. CCW), while reversing $m_{\mathrm{z},init}$ dictates whether the IFE-field polarity is antiparallel or parallel to the initial state. The lowest nucleation threshold is obtained when both conditions are favorable: the IFE field chirality matches the intrinsic DMI chirality and its polarity opposes $m_{\mathrm{z},init}$ (**Supplementary Figure S6**). Reversing only the DMI sign increases the threshold by approximately $15\mathrm{pJ}$ across the investigated pulse durations (**Supplementary Figure S7**), while only the favorable DMI sign enables nucleation for pulse durations as short as 20 ps. In contrast, setting the IFE core polarity parallel to $m_{\mathrm{z},init}$ requires substantially higher fluences and longer pulse durations. Strikingly, when both chirality and polarity are mismatched, the optical pulse can no longer drive the standard skyrmion-writing pathway and instead stabilizes a higher-order topological state: a skyrmionium (**Figure 3k** and **Supplementary Figure S6**). Symmetrically, under LHCP excitation, CW remains favored over CCW, while the favored initial magnetization is inverted from $m_{\mathrm{z},init} = -1$ to $+1$.

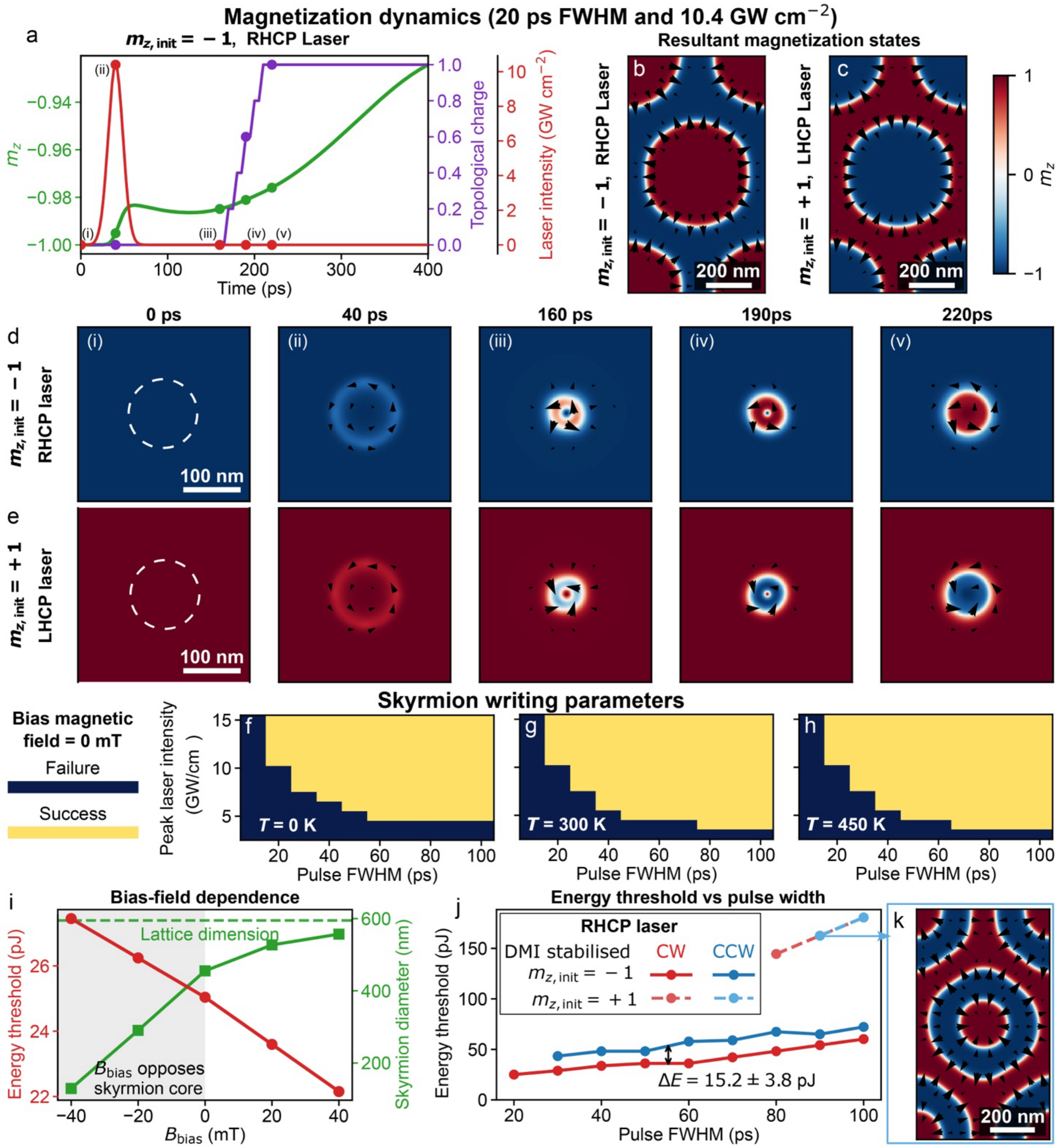

**Figure 3: Micromagnetic writing dynamics, parameter space, and multi-state switching. a)** Time evolution of the domain-averaged out-of-plane magnetization $m_z$ (green), normalized topological charge $Q$ (purple, calculated via the Berg–Lüscher method and normalized per skyrmion), and laser intensity profile (red) for an initial state $m_{z,init} = -1$ under RHCP illumination ($20$ ps FWHM, $10.4\ \mathrm{GW/cm^2}$). Points (i–v) mark key timestamps corresponding to the spatial snapshots in (d). **b,c)** Final relaxed magnetization states of the top magnetic layer in a single unit cell showing stable clockwise (CW) Néel skyrmions with core pointing up for $m_{z,init} = -1$ after an RHCP pulse (b) and core pointing down for $m_{z,init} = +1$ after an LHCP pulse (c). **d,e)** Time-resolved spatial snapshots of the top magnetic layer showing out-of-plane magnetization $m_z$ (color map) and in-plane components $(m_x, m_y)$ (arrows) at timestamps (i) $0\ \mathrm{ps}$, (ii) $40\ \mathrm{ps}$, (iii) $160\ \mathrm{ps}$, (iv) $190\ \mathrm{ps}$, and (v) $220\ \mathrm{ps}$ for (d) RHCP excitation ($m_{z,init} = -1$) and (e) LHCP excitation ($m_{z,init} = +1$), illustrating the transient Bloch-to-Néel pathway. **f–h)** Skyrmion writing phase diagrams (peak laser intensity vs. pulse FWHM; yellow = success, blue = failure) at zero bias field for temperatures of (f) $0\ \mathrm{K}$, (g) $300\ \mathrm{K}$, and (h) $450\ \mathrm{K}$. **i)** Nucleation energy threshold (red, left axis; integrated over the skyrmion nucleation area) and relaxed skyrmion diameter (green, right axis) as a function of the external bias field $\mathrm{B_{bias}}$ (the gray shaded area indicates the regime where $\mathrm{B_{bias}}$ opposes the skyrmion core; dashed green line denotes the lattice parameter). **j)** Nucleation energy threshold as a function of pulse FWHM under fixed RHCP illumination for four configurations varying the stack DMI and initial magnetization $m_{z,init}$. **k)** Final relaxed skyrmionium obtained when both the IFE-field chirality and polarity are mismatched with the preferred skyrmion state of the magnetic stack.

Beyond deterministic writing, a crucial capability of this magneto-plasmonic platform is that the established magnetic textures can be reconfigured or completely erased on demand by exploiting pulses of opposite optical helicity (**Figure 4**). Depending on the spatial confinement governed by an external bias field, a second optical pulse either achieves complete field-driven erasure of the skyrmion or converts it into a stable skyrmionium.

Under a confining perpendicular bias field ($\mathrm{B_{bias}} = -50\ \mathrm{mT}$) applied parallel to the initial background magnetization ($m_{z,init} = -1$), the written skyrmion is laterally constrained, preventing excessive domain expansion during the initial relaxation phase (**Figure 4a**). Following skyrmion nucleation by an RHCP pulse ($20\ \mathrm{ps}$ FWHM, $15\ \mathrm{GW/cm^2}$) and relaxation into the stable state ($Q = 1$, **Figure 4b**), a second laser pulse with reversed circular polarization (LHCP, $100\ \mathrm{ps}$ FWHM, $15\ \mathrm{GW/cm^2}$) is applied. Erasure requires a higher pulse energy than nucleation, here implemented through the longer pulse duration of the second excitation. Under the action of this reversed IFE field, the topological charge abruptly drops from $Q = 1$ to $Q = 0$ (**Figure 4a**). Upon subsequent energy dissipation, the multilayer fully recovers its pristine, uniformly magnetized single-domain state ($m_z = -1.00$, **Figure 4c**).

The microscopic erasure dynamics revealed in **Figure 4d** confirm that annihilation does not proceed through a simple radial contraction. Instead, the reversed IFE field initially reorganizes the pre-existing spin texture, driving the nucleation of an oppositely oriented central core surrounded by a transient ring-like perimeter ($200 - 310 \text{ ps}$). As the excitation subsides ($682 \text{ ps}$), this destabilized concentric pattern deforms and breaks into localized spin-wave fluctuations that fully vanish during relaxation ($890 \text{ ps}$), restoring the uniform ferromagnetic ground state across all layers without residual topological ($Q = 0$, **Figure 4a-b**).

In striking contrast, when the write-and-erase sequence is executed in the absence of a bias field ($B_{bias} = 0 \text{ mT}$), a completely different topological pathway is unlocked (**Figures 4e–h**). Without the confining bias field, the skyrmion nucleated by the first RHCP pulse expands substantially during the initial relaxation stage (**Figures 4e,f**). When the second LHCP pulse arrives, its localized IFE field acts primarily on the broad inner region of the expanded skyrmion rather than covering its outer perimeter. Consequently, the second pulse reverses the magnetization within the core interior without annihilating the outer domain wall (**Figures 4g,h**). Snapshots during the second pulse dynamics ($0$ to $372 \text{ ps}$, **Figure 4h**) show the progressive formation of an inner core of opposite polarity encased within the pre-existing outer boundary. Following relaxation, this concentric, target-like spin texture stabilizes as a robust **skyrmionium** ($Q = 0$, **Figure 4e-g**).

This demonstrates that by modulating the spatial confinement of the initial skyrmion via a weak external bias magnetic field, the same optical excitation pulse can be directed either to achieve complete, field-driven topological erasure or to deterministically synthesize multi-ring topological quasiparticles.

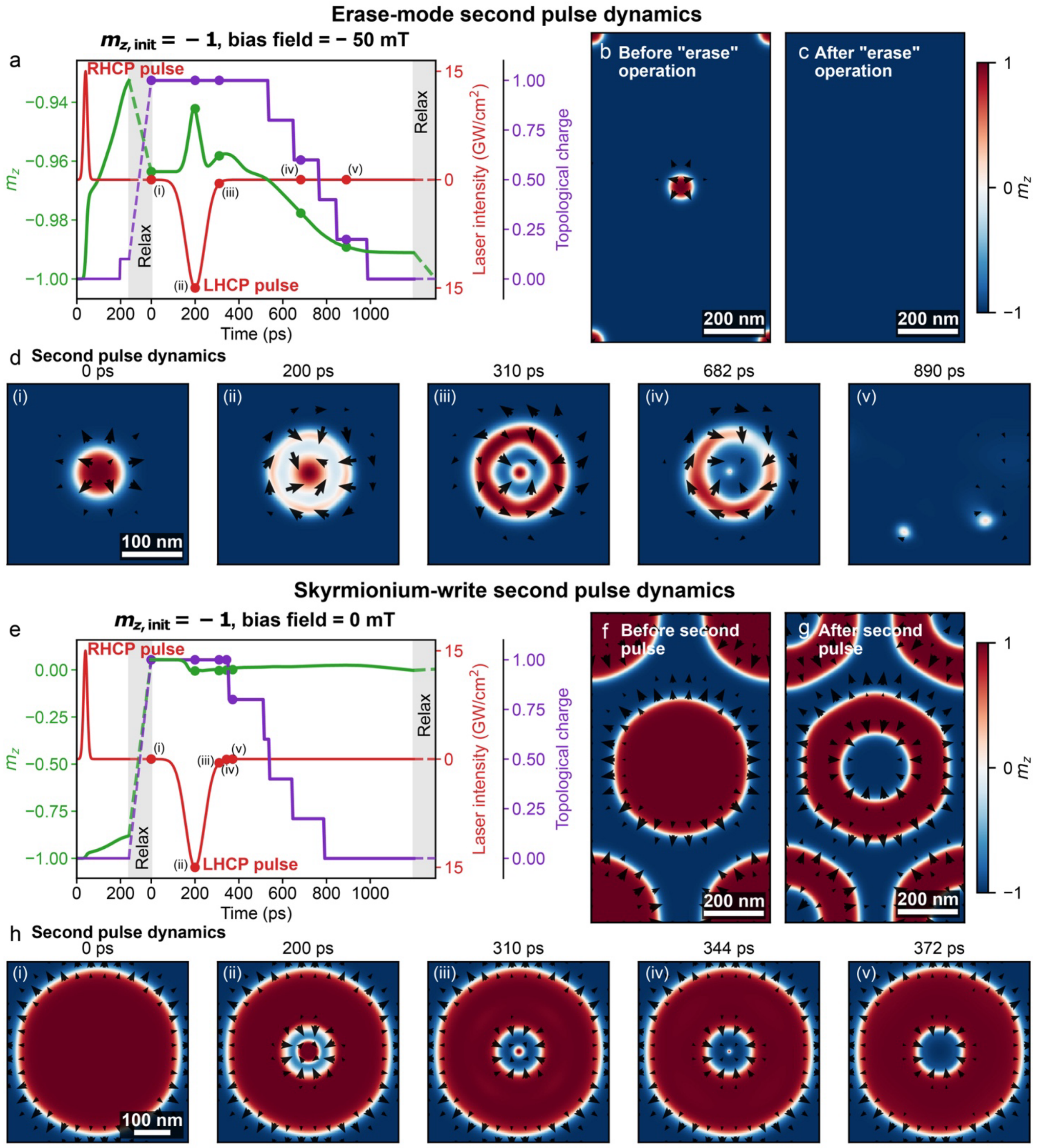

**Figure 4: Reversible all-optical skyrmion erasure and skyrmionium formation via helicity-reversed second pulse. a–d)** All-optical skyrmion erasure under a confining bias magnetic field (Erase-mode second pulse under a confining bias field, $\mathrm{B_{bias}} = -50\ \mathrm{mT}$). **a)** Time evolution of domain-averaged $\mathrm{m_z}$ (green), normalized topological charge $\mathrm{Q}$ (purple), and laser intensity (red) during a consecutive write-and-erase sequence: RHCP write pulse ($20\ \mathrm{ps}$ FWHM, $15\ \mathrm{GW/cm^2}$), intermediate relaxation ($\mathrm{Q} = 1$), LHCP erase pulse ($100\ \mathrm{ps}$ FWHM, $15\ \mathrm{GW/cm^2}$), and final relaxation ($Q = 0$). **b,c)** Relaxed spatial distribution of $m_z$ in a single unit cell (b) before erasure (stable written skyrmion) and (c) after erasure, confirming complete recovery of the uniform ground state ($m_z = -1$). **d)** Time-resolved snapshots of

the top layer during the LHCP pulse showing $m_z$ (color map) and in-plane components (arrows) at 0, 200, 310, 682, and $890\,\mathrm{ps}$, showing texture deformation followed by topological collapse. **e–h)** Skyrmionium writing (Skyrmionium-write second pulse in the absence of a bias field, $\mathrm{B_{bias}} = 0\,\mathrm{mT}$). **e)** Time evolution of $\mathrm{m_z}$, normalized $\mathrm{Q}$, and laser intensity for the same RHCP–LHCP pulse sequence at zero field. **f,g)** Relaxed magnetization states (f) before the second pulse (expanded skyrmion) and (g) after the second pulse, showing the formation of a stable skyrmionium with two concentric domain walls. **h)** Time-resolved snapshots during the second LHCP pulse at 0, 200, 310, 344, and $372\,\mathrm{ps}$, tracking the reversal of the central core within the expanded skyrmion into a skyrmionium.

Finally, regarding the practical implementation of this platform, it is important to address the laser fluence requirements. The peak intensity of $10.4\,\mathrm{GW/cm^2}$ combined with picosecond pulse durations yields an integrated energy that might approach the optical damage threshold of standard plasmonic nanostructures. However, extensive experimental studies have consistently demonstrated that the effective IFE-induced magnetic fields in plasmonic and metallic systems can be up to three orders of magnitude stronger than those predicted by standard classical hydrodynamic models.[36] Consequently, the actual laser fluences required to achieve the necessary magnetic field threshold in physical experiments may be significantly lower, in line with prior experimentally demonstrated enhancements,[36] placing the operational window safely within the non-destructive regime of the metasurface.

In summary, we have demonstrated a deterministic, field-driven platform for the all-optical writing and erasing of magnetic skyrmion crystals by interfacing a plasmonic metasurface with a chiral magnetic multilayer. By exploiting surface lattice resonances in a periodic array of Ag nanostructures, circularly polarized light generates strongly enhanced local optical spin densities that drive intense circulating drift photocurrents via the inverse Faraday effect. This mechanism delivers synchronized, localized out-of-plane magnetic field pulses directly into the adjacent magnetic medium, bypassing the stochasticity, structural damage, and multi-domain fragmentation inherent to conventional thermally driven nucleation methods.

Our micromagnetic simulations detail the microscopic dynamics governing this topological imprinting, where the optomagnetic field initiates a transient Bloch-like spin texture whose circulation handedness is strictly dictated by the optical helicity, before relaxing into a stable Néel skyrmion lattice stabilized by interfacial Dzyaloshinskii–Moriya interaction which determines its chirality. We establish that thermal fluctuations assist this switching process by lowering the required optical energy barrier without compromising determinism, while external bias fields and material chirality matching offer versatile knobs to modulate the nucleation threshold and final skyrmion dimensions.

Crucially, this approach provides routes to reconfigurable multi-state functionality. Simply reversing the circular polarization of the laser pulse enables complete topological erasure under a confining bias field, or alternatively, drives the deterministic conversion of an expanded skyrmion into a stable skyrmionium in the absence of a bias field. By demonstrating reversible, field-driven topological writing, erasing, and multi-state switching on picosecond timescales, this magneto-plasmonic platform establishes a foundation for topological data storage and unconventional & neuromorphic computing architectures.

**Availability of data and material**

The materials, data and code underlying the results presented in this paper are not publicly available at this time but may be obtained from the authors upon reasonable request.

**Acknowledgements**

This work is supported by the ERC grant FemtoMagnet (grant no. 101087709). We acknowledge the financial support from the Agence Nationale de la Recherche (ANR-20-CE09-0031-01, ANR-22-CE09-0027-04 and ANR-23-ERCC-0005), the Institut de Physique du CNRS (Tremplin@INP 2020). JCG was supported by the Royal Academy of Engineering Research Fellowship RF2122-21-363, the EPSRC ECR International Collaboration Grant EP/Y003276/1., the ERC Starting Grant MORPHON, and the Imperial College London President's Excellence Fund for Frontier Research. This work was supported by a France 2030 Government grant managed by the French National Research Agency (ANR) PEPR SPIN CHIREX (Grant No. ANR-22-EXSP-0002). Financial support was provided by the European Union. Views and opinions expressed are however those of the authors only and do not necessarily reflect those of the European Union or the European Research Executive Agency (REA). Neither the European Union nor the granting authority can be held responsible for them. T.d.C.S.C.G. is an ERA fellow. Computational resources have been provided by the Consortium des Équipements de Calcul Intensif (CÉCI), funded by the Fonds de la Recherche Scientifique de Belgique (F.R.S.-FNRS) under Grant No. 2.5020.11 and by the Walloon Region.

**Author contributions**

M.M., M.S.-P., N.R. and J.C.G. conceived of and designed the research. X.Y. performed the electromagnetic simulations with contributions from Y.M., M.S.-P. and C.H. T.d.C.S.C.G. carried out the micromagnetic simulations, with T.Z. and J.C.G. co-developing the micromagnetic simulation framework and J.C.G. co-running simulations. X.Y., T.d.C.S.C.G., T.Z., J.C.G., N.R., M.S.-P. and M.M. analyzed the data. M.M. prepared the original draft with contributions from all authors.

**Competing interests**

The authors declare no competing interests.

**Correspondence and requests for materials** should be addressed to Mathieu Mivelle.